\documentclass[12pt,a4paper]{article}
\usepackage{epsfig,graphics,amsmath,amssymb,here}

\makeatletter
\newif\if@preliminary
\@preliminaryfalse
\def\preliminary{\@preliminarytrue}
\def\preliminary{\@preliminaryfalse}
\def\preprintno#1{\def\@preprintno{#1}}
\def\address#1{\def\@address{#1}}

\def\abstract#1{\def\@abstract{#1}}
\renewcommand\abstractname{ABSTRACT}
\newlength\preprintnoskip
\newlength\abstractwidth
\@titlepagetrue
\renewcommand\maketitle{\begin{titlepage}%
  \let\footnotesize\small
  \hfill\parbox{\preprintnoskip}{%
  \begin{flushright}\@preprintno\end{flushright}}\hspace*{1cm}
  \vskip 60\p@
  \begin{center}%
    {\Large\bf\boldmath \@title \par}\vskip 1cm%
    {\sc\@author \par}\vskip 3mm%
    {\@address \par}%
    \if@preliminary
      \vskip 2cm {\large\sf PRELIMINARY DRAFT \par \@date}%
    \fi
  \end{center}\par
  \@thanks
  \vfill
  \begin{center}%
    \parbox{\abstractwidth}{\centerline{\abstractname}%
    \vskip 3mm%
    \@abstract}
  \end{center}
  \end{titlepage}%
  \setcounter{footnote}{0}%
  \let\thanks\relax\let\maketitle\relax
  \gdef\@thanks{}\gdef\@author{}\gdef\@address{}%
  \gdef\@title{}\gdef\@abstract{}\gdef\@preprintno{}
}%
\def\@citex[#1]#2{\if@filesw\immediate\write\@auxout{\string\citation{#2}}\fi
  \def\@citea{}\@cite{\@for\@citeb:=#2\do
    {\@citea\def\@citea{,\penalty\@m}\@ifundefined
       {b@\@citeb}{{\bf ?}\@warning
       {Citation `\@citeb' on page \thepage \space undefined}}%
\hbox{\csname b@\@citeb\endcsname}}}{#1}}
\def\citerange{\@ifnextchar [{\@tempswatrue\@citexr}{\@tempswafalse\@citexr[]}}
\def\@citexr[#1]#2{\if@filesw\immediate\write\@auxout{\string\citation{#2}}\fi
  \def\@citea{}\@cite{\@for\@citeb:=#2\do
    {\@citea\def\@citea{--\penalty\@m}\@ifundefined
       {b@\@citeb}{{\bf ?}\@warning
       {Citation `\@citeb' on page \thepage \space undefined}}%
\hbox{\csname b@\@citeb\endcsname}}}{#1}}
\long\def\@makecaption#1#2{%
  \vskip\abovecaptionskip
  \sbox\@tempboxa{#1: \emph{#2}}%
  \ifdim \wd\@tempboxa >\hsize
    #1: \emph{#2}\par
  \else
    \hbox to\hsize{\hfil\box\@tempboxa\hfil}%
  \fi
  \vskip\belowcaptionskip}
\def\fmslash{\@ifnextchar[{\fmsl@sh}{\fmsl@sh[0mu]}}
\def\fmsl@sh[#1]#2{%
  \mathchoice
    {\@fmsl@sh\displaystyle{#1}{#2}}%
    {\@fmsl@sh\textstyle{#1}{#2}}%
    {\@fmsl@sh\scriptstyle{#1}{#2}}%
    {\@fmsl@sh\scriptscriptstyle{#1}{#2}}}
\def\@fmsl@sh#1#2#3{\m@th\ooalign{$\hfil#1\mkern#2/\hfil$\crcr$#1#3$}}
\makeatother

\def\fmfL(#1,#2,#3)#4{\put(#1,#2){\makebox(0,0)[#3]{#4}}}

\newcommand\OBS{\mbox{$\cal{O}$}}
\newcommand\PRO{\mbox{$\wp$}}
\newcommand\ZJ{\mbox{Z $\rightarrow$ 4 jets}}

\newcommand\ZQG{\mbox{$Z \rightarrow q \bar{q} G G$}}
\newcommand\ZBG{\mbox{$Z \rightarrow b \bar{b} G G$}}
\newcommand\ZCG{\mbox{$Z \rightarrow c \bar{c} G G$}}
\newcommand\ycut{\mbox{$y_{cut}$}}
\newcommand\hg{\mbox{$\widehat{\kappa}$}}
\newcommand\dhg{\mbox{$\delta\widehat{\kappa}$}}
\begin{document}
\preliminary        
\baselineskip20pt   
\preprintno{HD--THEP 99--41\\[0.5\baselineskip] September 1999}
\title{%
 THE CP-VIOLATING TRIPLE GLUON INTERACTION IN Z
$\rightarrow$ 4 JETS 
\footnote{Supported by German Bundesministerium f\"ur Bildung und
Forschung (BMBF),\\ Contract Nr.~05~7HD~91~P(0), and by the
Landesgraduiertenf\"orderung}}
\author{%
 O.~Nachtmann
 and C.~Schwanenberger
}
\address{%
 Institut f\"ur Theoretische Physik, Universit\"at Heidelberg,
 Philosophenweg 16\\
 D--69120 Heidelberg, Germany\\
{\footnotesize O.Nachtmann@thphys.uni-heidelberg.de, C.Schwanenberger@thphys.uni-heidelberg.de}
}
\abstract{%
We analyse CP-violating effects in Z $\rightarrow$ 4 jet 
decays, assuming the
presence of a CP-violating effective triple gluon coupling. We discuss
the influence of this coupling on the decay width.
Furthermore, we analyse different CP-odd observables and propose strategies
of a direct 
search for such a CP-violating $GGG$ coupling. The present data of LEP 1
should give significant information on the coupling. 
}
\maketitle


\setcounter{footnote}{3}
\section{Introduction}

In electron-positron collider experiments at LEP and SLC, a large number of
Z bosons has been collected so that the detailed study of the decays of the
Z boson has been 
made possible \cite{lepwg}. An
interesting topic is the test of CP symmetry in
such Z decays.  There is already a number of theoretical 
(\cite{othertheo1}-\cite{over} and references therein) and experimental
\cite{opaltautau}-\cite{opal}
studies of this subject. In the present paper we will study a
flavour-diagonal Z decay where
CP-violating effects within the
Standard Model (SM) are estimated to be very small \cite{zdecay}. Thus,
looking for CP violation in such Z decays
means looking for new physics beyond the SM. 

For a model-independent systematic analysis of CP violation in Z decays
we use the effective Lagrangian approach as described in
\cite{zdecay,xsec}. Here we consider the 4 jet decays of the
Z boson. In \cite{width,higgs,paper} the effects of CP violating couplings
involving heavy quarks were studied for 3 and 4 jet decays of the Z
boson. On the other hand the 4 jet decays offer also the possibility to
study the CP violating triple gluon coupling
which was 
listed in \cite{morozov,buchmueller} and was investigated in
\cite{tripgl1}-- \cite{tripgl3}.
The following three subprocesses contribute to the 4 jet decay:
\begin{eqnarray}
\label{proc4}\nonumber
  \!\!\!\!\!\!\!\!\!\!\!
  e^+\,(p_+,v) \; e^-\,(p_-,w)
  \rightarrow  Z\,(p,j) \rightarrow 
  q'\,(k_-,s,B)\; \bar{q'}\,(k_+,r,A)\;
  G\,(k_1,\kappa_1,a)\; G\,(k_2,\kappa_2,b) \, ,\!\!\!\!\!\!\!\!\!\!\!& &\\
   & & 
\end{eqnarray}
\begin{eqnarray}
\label{proc5}\nonumber
  \!\!\!\!\!\!\!\!\!\!
  e^+\,(p_+,v) \; e^-\,(p_-,w)
  \rightarrow  Z\,(p,j) \rightarrow 
  q'\,(k_-,s,B)\; \bar{q'}\,(k_+,r,A)\;
  q'\,(q_-,u,D)\; \bar{q'}\,(q_+,t,C) \, ,\!\!\!\!\!\!\!\!\!\!\!& &\\
   & & 
\end{eqnarray}
\begin{eqnarray}
\label{proc6}\nonumber
  \!\!\!\!\!\!\!\!\!\!
  e^+\,(p_+,v) \; e^-\,(p_-,w)
  \rightarrow  Z\,(p,j) \rightarrow 
  q'\,(k_-,s,B)\; \bar{q'}\,(k_+,r,A)\;
  q\,(q_-,u,D)\; \bar{q}\,(q_+,t,C) \, ,\!\!\!\!\!\!\!\!\!\!\!& &\\
 (\:q\neq q'\:) \, , \;\;\;
\end{eqnarray}
where $q$ and $q'$ denote quarks with flavour $q,q'= u, d, s, c, b$. 
We will always assume unpolarized $e^+$, $e^-$ beams. Only for process
(\ref{proc4}) the CP-violating $GGG$ coupling comes into play. We will show
the results for this process alone as well as the
results for the sum of them. In the experiments, of course, only the sum of
the three processes can be observed easily. 

In chapter 2 we explain the theoretical framework of our
computations. Next, in chapter 3, we analyse the anomalous coupling for
partons in the final state. First, we discuss anomalous contributions to the
decay width. Then, we investigate different CP-odd tensor observables
as in \cite{paper}
and calculate their sensitivities to the anomalous coupling. In order to
find out 
how ``good'' for the measurement of the new coupling our observables are, we
compare them to the optimal observable. In chapter 4 we study the optimal
observable in a realistic scenario for an
experimental analysis. Our conclusions can be found in chapter 5.

\section{Effective Lagrangian Approach}

\begin{figure}[ht]
  \begin{center} \epsfig{file=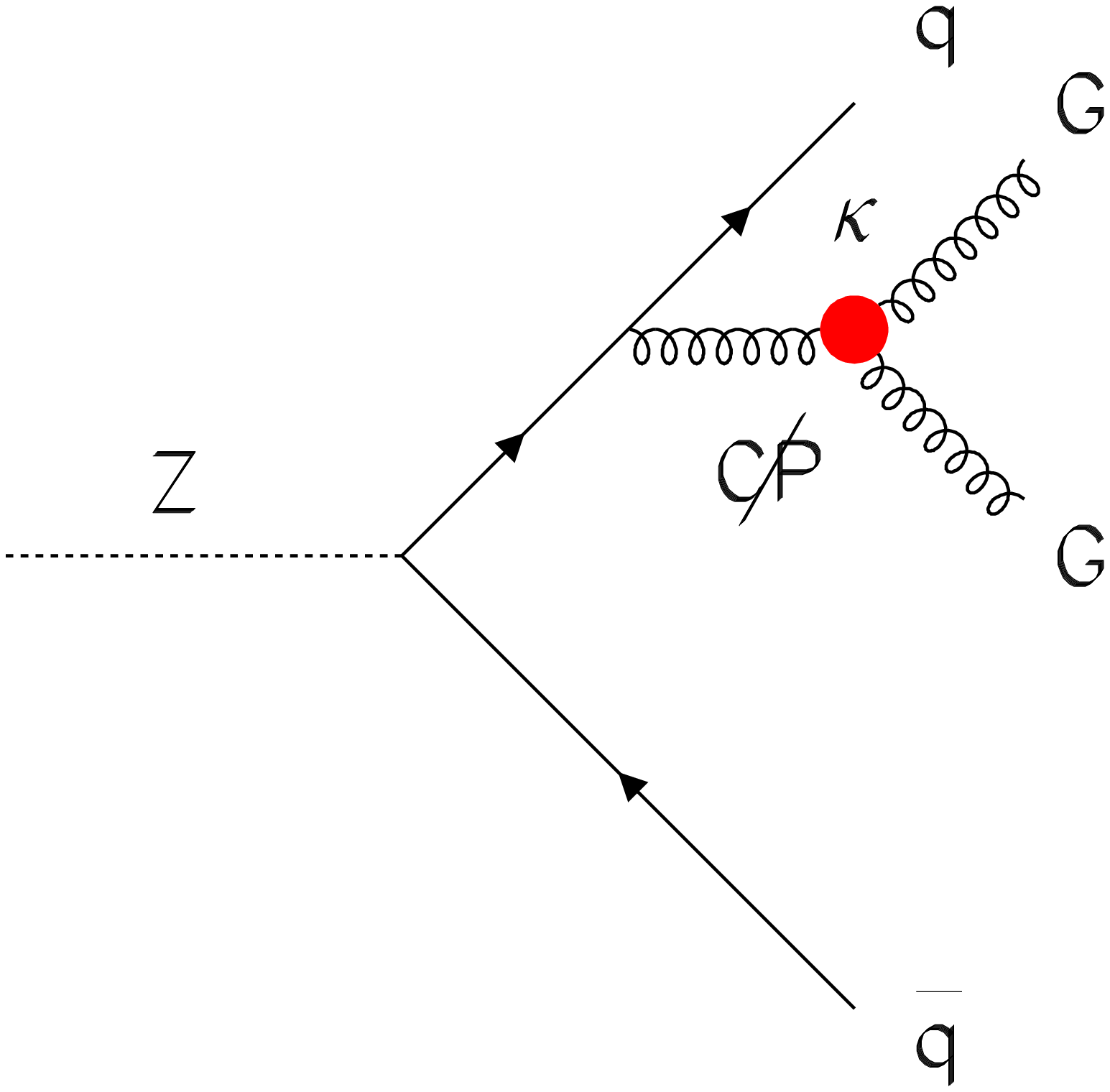,width=0.5\hsize}
    \parbox{0.9\textwidth}
{\caption{\it{
Diagram for \ZQG\ with the CP-violating vertex (\ref{lcp}).
    }}\label{fig:vert3g}
}  \end{center}
\end{figure}

For a model independent study of CP violation in 4 jet
decays of the Z boson we use the effective Lagrangian approach as explained
in \cite{zdecay}. We add to the SM Lagrangian ${\cal L}_{SM}$ the following 
CP-violating term which contains a mass dimension $d \leq
6$ local operator involving 3 gluons:

\begin{equation}
  \label{lcp} {\cal L}_{CP}(x) = 
  i \: \kappa \: Sp \{ \: G_{\alpha\beta}(x) \: G_{\mu\nu}(x) \:
  G_{\rho\sigma}(x) \: \} \: \epsilon^{\alpha\beta\mu\rho} \: g^{\nu\sigma}
  \;\;,
\end{equation}
where $G^a_{\mu\nu}(x)$ represents the field strength tensor of the gluon
and $\epsilon^{\alpha\beta\mu\rho}$ is the totally antisymmetric tensor
with $\epsilon_{0123} = + 1$.
A typical process where we find the corresponding vertex following from ${\cal L}_{CP}$ is shown 
in figure~\ref{fig:vert3g}.

We define a dimensionless coupling constant \hg\ using the Z
mass as the scale parameter by
\begin{equation}
\label{kappa}
        \kappa = \frac{g_s}{m_Z^2}\; \widehat{\kappa} \;.
\end{equation}
Here $g_s$ is the gauge coupling constant of QCD.
For numerical calculations we set $m_Z=91.187\:$GeV and $\alpha_s \equiv
g_s^2/4\pi =0.118$ since we consider a process at the Z mass scale
\cite{pdg}. Our calculations are carried out in
leading order of the 
CP-violating coupling of
${\cal L}_{CP}$ and the SM couplings. All quark masses are
neglected.\footnote{For further details of the calculation we refer to
\cite{dok}.} 

\section{Study of the CP-violating coupling for partons in the final state}
\label{sec:partons3g}

In this chapter we discuss an ideal experiment where one is able to
flavour-tag the partons and measure their momenta. We present a study
of the CP-violating coupling for process (\ref{proc4}) with $q'=u,c$ and 
$q'=d,s,b$ separately and for the sum of the processes (\ref{proc4}) --
(\ref{proc6}). We have computed the differential and integrated decay rates using FORM
\cite{form} and M \cite{M} for the analytic and VEGAS \cite{vegas} 
for the numerical calculation. We write the squared matrix element for the
process (\ref{proc4}) with final state \PRO,
\begin{equation}
\label{procstuzsmcp}
\PRO = {u \bar{u} G G},\; {d \bar{d} G G},\;  {s \bar{s} G G},\; {c
  \bar{c} G G},\; {b \bar{b} G G} \;,
\end{equation}
in the form:
\begin{equation}
\label{megzerl}
  R(\phi)^{(\PRO)} = S_0(\phi)^{(\PRO)} + \hg S_1(\phi)^{(\PRO)} + \hg^2 S_2(\phi)^{(\PRO)} \;.
\end{equation}
Here $\phi$ stands collectively for the phase space variables,
$S_0$ denotes the SM part.
For the processes (\ref{proc5}), (\ref{proc6}) with final state \PRO,
\begin{eqnarray}
\nonumber
\label{procstuzsm}
\PRO &=& {u \bar{u} u \bar{u}},\; {d \bar{d} d \bar{d}},\; {s \bar{s} s
  \bar{s}},\; {c \bar{c} c \bar{c}},\; {b \bar{b} b \bar{b}},\;\nonumber \\
& & {u \bar{u} c \bar{c}},\; \nonumber \\
& & {u \bar{u} d \bar{d}},\; {u \bar{u} s \bar{s}},\; {u \bar{u} b \bar{b}},\; 
{c \bar{c} d \bar{d}},\; {c \bar{c} s \bar{s}},\; {c \bar{c} b \bar{b}},\;
\nonumber \\ 
& & {d \bar{d} s \bar{s}},\; {d \bar{d} b \bar{b}},\; {s \bar{s} b \bar{b}} \;,
\end{eqnarray}
the matrix element contains only a SM part:
\begin{equation}
\label{megzerl2}
  R(\phi)^{(\PRO)} = S_0(\phi)^{(\PRO)} \;.
\end{equation}

The definition of a 4 jet sample requires the introduction of
resolution cuts. We use JADE cuts \cite{jade} requiring
\begin{equation}
        y_{ij} = \frac{2\, E_iE_j\,(1-\cos \vartheta_{ij})}{m_Z^2} >
        y_{cut} \;,
\label{jade}
\end{equation}
with $\vartheta_{ij}$ the angle between the momentum directions of any two
partons ($i \neq j$) and $E_i$, $E_j$ their energies in the Z rest
system.
The expectation value of an observable ${\cal{O}}(\phi)$ is then defined as
\begin{equation}
   <{\cal{O}}> = \frac{\int {\cal{O}} (\phi) \; R(\phi) \; d \phi}{\int
     R(\phi) \; d \phi} \, . 
\label{erwwobsallg}
\end{equation}

\subsection{Anomalous contributions to the decay widths}

\begin{figure}[H]
  \begin{center} \epsfig{file=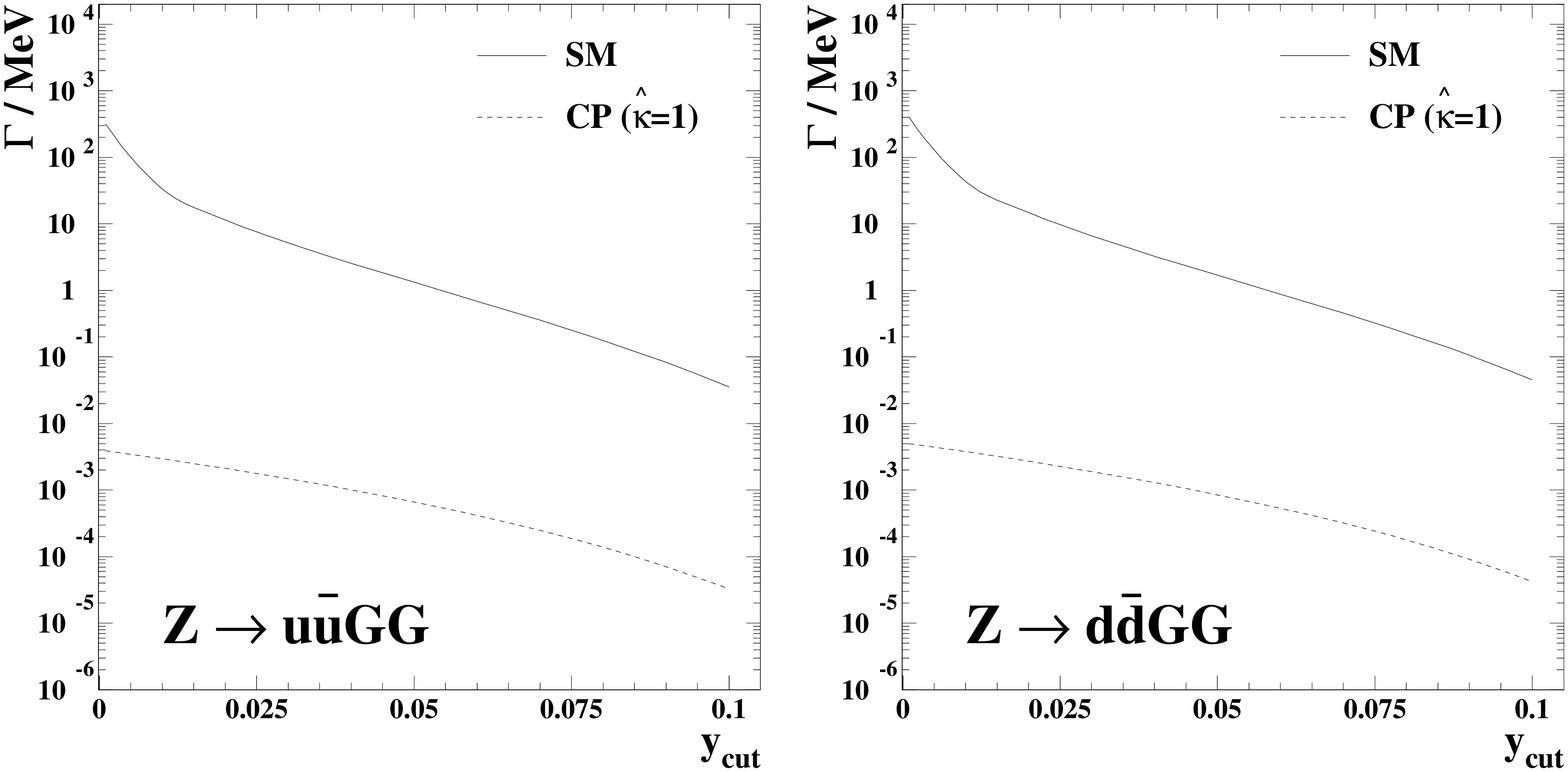,width=\hsize}\vspace{-7.5cm}
    \parbox{0.9\textwidth}{\caption{\it{
The decay width for different subprocesses as function of the jet
resolution parameter $y_{cut}$ (\ref{jade}). The results for $Z
  \rightarrow c \bar{c} G G $ ($s \bar{s} G G$, $b \bar{b} G G$) 
are identical to those of {$Z \rightarrow u \bar{u} G G $} ($d \bar{d} G
    G$).
    }}\label{fig:widsubhg}
}  \end{center}
\end{figure}

The solid curves in figure~\ref{fig:widsubhg} show the results of our
calculations for the SM decay widths $\Gamma^{SM}$ as function of the jet
resolution parameter 
$y_{cut}$ for process (\ref{proc4}) with the different final states \PRO\
of (\ref{procstuzsmcp}).
To check our calculations we computed $\Gamma^{SM}$ also with the program
COMPHEP \cite{comphep} and found --- within numerical errors --- complete
agreement.

\begin{figure}[H]
  \begin{center} \epsfig{file=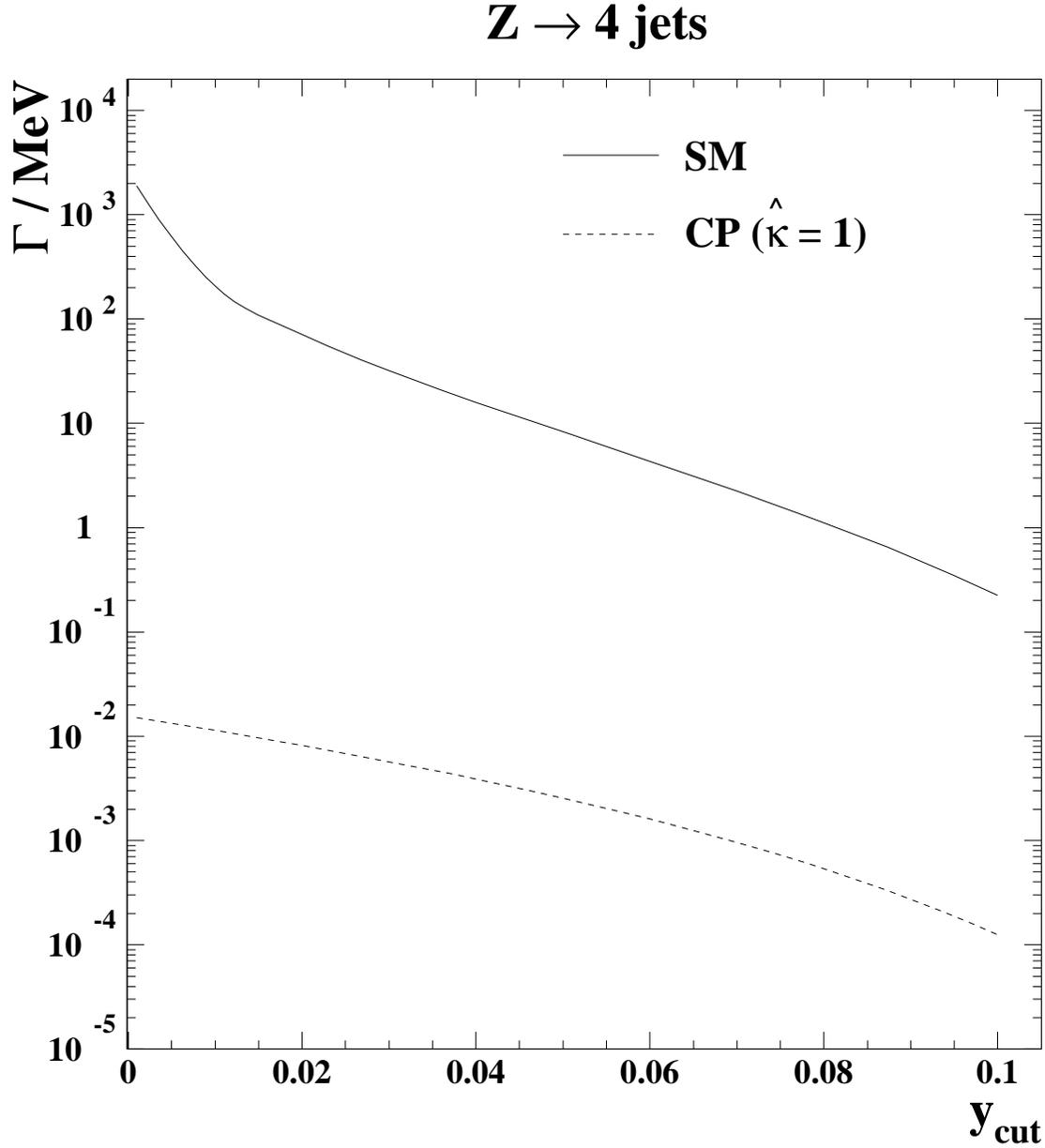,width=\hsize}
    \parbox{0.9\textwidth}{\caption{\it{
The 4 jet decay width (sum of processes (\ref{proc4}) -- (\ref{proc6}) ) as
    function of the jet resolution parameter $y_{cut}$ (\ref{jade}).
    }}\label{fig:widhg}
}  \end{center}
\end{figure}

As the decay width is a CP-even observable the contribution of the
CP-violating interaction to it adds incoherently to the SM one:
\begin{equation}
        \Gamma = \Gamma^{SM} + \Delta\Gamma^{CP} \;,
\end{equation}
with $\Delta\Gamma^{CP}$ being quadratic in the
new coupling. In figure~\ref{fig:widsubhg}
the dashed curves represent $\Delta\Gamma^{CP}$ as function of $y_{cut}$
assuming \mbox{$\hg=1$}. 

In Figure~\ref{fig:widhg}, we compare 
for the sum of the processes (\ref{proc4}) -- (\ref{proc6}) $\Gamma^{SM}$ and 
$\Delta\Gamma^{CP}$ assuming \mbox{$\hg=1$}.
$\Delta\Gamma^{CP}$ is only a correction of less than a per mille to
$\Gamma^{SM}$. 
Thus, considering the theoretical uncertainties in the SM 4 jet decay rate,
a determination of the new coupling
by measuring the decay width alone does not look promising.
 
\subsection{CP-odd observables for \ZBG\ and \ZCG}
\label{sec:partonscb3g}

Tagging of jets according to quark flavour or gluon is a difficult
experimental task. What can be done routinely now is tagging of heavy quark
$b$ and $c$ jets. Thus we study in this section the information obtainable
on $\kappa$ (\ref{lcp}) from reaction (\ref{proc4}) with $q'=c$ and $b$.

Let us first consider   
CP-odd observables constructed from the
momentum directions of the $q$ and $\bar{q}$ quarks ($q = c, b$),
$\widehat{\bf k}_q={\bf k}_q/|{\bf k}_q|$  and $\widehat{\bf k}_{\bar{q}}={\bf k}_{\bar{q}}/|{\bf k}_{\bar{q}}|$
(cf. \cite{zdecay,xsec,bernnach2,hab,paper}):
\begin{equation}
        T_{ij}^{(n)} = (\widehat{\bf k}_{\bar{q}} - \widehat{\bf k}_q)_i \; (\widehat{\bf k}_{\bar{q}} \times
\widehat{\bf k}_q)_j \; |\widehat{\bf k}_{\bar{q}} \times \widehat{\bf k}_q|^{n-2} + (i \leftrightarrow j) \; ,
\label{ten}
\end{equation}
with $i$, $j$ the Cartesian vector indices in the Z rest system and
$n=1,2,3$.

The observables $T_{ij}^{(n)}$ transform as tensors. For unpolarized
$e^+e^-$ beams and our rotationally invariant cuts 
(\ref{jade}) 
their expectation values are then proportional to the Z
tensor polarization $S_{ij}$. Defining the positive $z$-axis in the $e^+$ beam
direction, we have
\begin{equation}
  (S_{ij}) = \frac{1}{6} \left( \begin{array}{ccc}
  -1 & 0 & 0 \\
  0 &-1 & 0 \\
  0 & 0 & 2    
\end{array} \right) \;.
\end{equation}
This shows that the components $T_{33}^{(n)}$ are the most
sensitive ones. 

Note that the tensor observables do {\em not} change their sign upon charge
misidentification ($\widehat{\bf k}_{\bar{q}} \leftrightarrow \widehat{\bf
k}_q$). Thus there is no need of charge identification in a measurement.
We have also investigated vector observables like in \cite{paper}, but found
them to be scarcely sensitive on the CP-violating coupling (\ref{lcp}).

We have computed the expectation values of the observables (\ref{ten}),
for different JADE cuts (\ref{jade}), as function of
\hg.
The expectation value of a CP-odd observable ${\cal{O}}$
has the following general form:

\begin{equation}
        <\!{\cal{O}}\!> \, = \, c \, \hg \;
        \frac{\Gamma^{SM}}{\Gamma} \;,
\end{equation}
where $\Gamma^{SM}$ and $\Gamma$ denote the
corresponding \ZJ\
decay widths in the SM and in the theory with SM plus CP-violating coupling,
respectively. 
In an experimental analysis $\Gamma^{SM}$ should be
taken from the theoretical calculation, $\Gamma$ and
$<\!{\cal{O}}\!>$ from the experimental measurement. The quantity
$<\!{\cal{O}}\!> \! \cdot \,\Gamma$ is then an observable
strictly linear in 
the anomalous coupling.

The relative statistical error $\delta{\widehat{\kappa}}$ to leading order in the anomalous
coupling 
in a
measurement of the coupling \hg\ using the observable ${\cal{O}}$ is given by:
\begin{equation}
\label{dhg}
        \dhg =
        \frac{\sqrt{<\!{\cal{O}}^2\!>_{SM}}}{|c|\sqrt{N}} \;,
\end{equation}
where $N$ is the number of events within cuts. 
A measure for the sensitivity of \OBS\ to \hg\ is $1/\dhg$.

In addition to the tensor observables (\ref{ten})
we study the
{\em optimal observable}, which has the largest possible statistical
signal-to-noise ratio \cite{opt,opt1,opt2}. Neglecting higher orders 
in the anomalous coupling the optimal
observable for measuring \hg\ is obtained from the differential
cross sections (\ref{megzerl}) and (\ref{megzerl2}), respectively, as
\begin{equation}
\label{opthg}
  O(\phi) = \frac{S_1(\phi)}{S_0(\phi)} \;.
\end{equation}
The expectation value has then the following form:
\begin{equation}
\label{evopthg}
        <\!O\!> = c \, \hg \;,
\end{equation}
with the coefficient
\begin{equation}
  c = \frac{1}{\int S_0 d\phi}
  \int \frac{S_1(\phi)}{S_0(\phi)}S_1(\phi)d\phi \;.
\end{equation}

%
\begin{figure}[H]
\vspace{-3cm}
  \begin{center} \epsfig{file=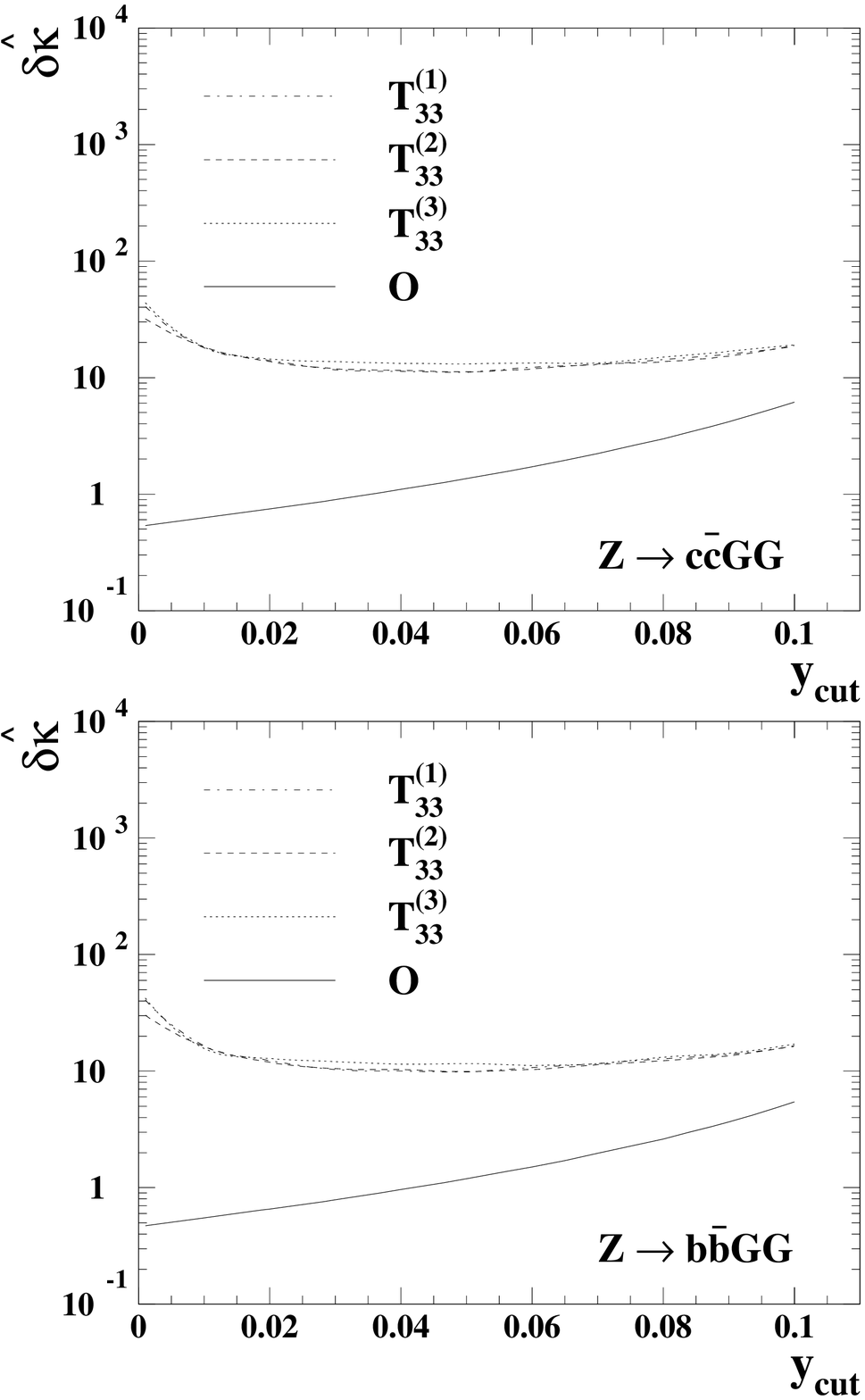,width=0.85\hsize}
    \parbox{0.9\textwidth}
{\caption{\it{
The inverse sensitivities of tensor and optimal observables to \hg\
obtainable in the subprocesses (\ref{proc4}) with $q'=c$ and $q'=b$
    as function of the jet resolution parameter $y_{cut}$
    (\ref{jade}) assuming (\ref{nevents3g}) for the number of events.
    }}\label{fig:obsqqgg}
}  \end{center}
\end{figure}

We have calculated the sensitivities to \hg\
for different tensor observables and the optimal
observable varying the jet resolution parameter \ycut\ .
We assume a total number of $5 \cdot 10^5$ 4 jet events from (\ref{proc4}
-- \ref{proc6}) for $y_{cut}=0.01$:
\begin{equation}
  \label{nevents3g}
  N(y_{cut}=0.01)=500000 \;.
\end{equation}
The number of events for other values
of $y_{cut}$ and for the various subprocesses is then calculated within
the SM. The total number of Z decays corresponding to (\ref{nevents3g}) is
$N_{tot} \cong 6 \cdot 10^6$.

In figure~\ref{fig:obsqqgg} we show the inverse sensitivities to the CP-odd
triple 
gluon coupling as calculated from (\ref{dhg}).
The
differences due to the different weight factors for tensor observables
$T_{33}^{(n)}$ ($n=1,2,3$) are only small but
all these observables have significantly lower sensitivities than the
optimal observable. Information on the spin of the final state partons in
(\ref{proc4}) -- 
(\ref{proc6}) is hardly available experimentally. Thus, we consider as
observables only the parton's energies and momenta. Then, we didn't find any
simple CP-odd observables with a significantly better sensitivity than those
of the tensor
observables $T_{33}^{(n)}$ ($n=1,2,3$). It is thus
of advantage to use the more complicated but much more sensitive optimal
observable for the experimental analysis of the CP-odd coupling.

In tables~\ref{tab:optparthg} -- \ref{tab:opthgbbgg}
in appendix~\ref{sec:numvalues} we list the coefficient of the expectation
value (\ref{evopthg}) for the optimal observable (\ref{opthg})
for different values of the jet resolution parameter
$y_{cut}$ (\ref{jade}) for the sum of the reactions (\ref{proc4}) --
(\ref{proc6}) and for the reaction (\ref{proc4}) with $q'=c$ and $b$,
respectively.

\section{CP-violating observables for untagged jets}
\label{sec:jets3g}

In this section we consider an experimental analysis of untagged 
jets which are ordered according to the magnitude of
their momenta ${\bf q}_i$, $i=1,2,3,4$ : \footnote{This we called analysis 4
in \cite{paper}. Further details can be found therein.}
\begin{equation}
  |{\bf q}_1| \geq |{\bf q}_2| \geq |{\bf q}_3| \geq |{\bf q}_4| \;.
\label{momorder}
\end{equation}
Thus, all processes (\ref{proc4}) -- (\ref{proc6}) are lumped together here.

%
\begin{figure}[H]
  \begin{center} 
    \epsfig{bbllx=0bp,bblly=350bp,bburx=500bp,bbury=760bp,width=0.9\hsize,file=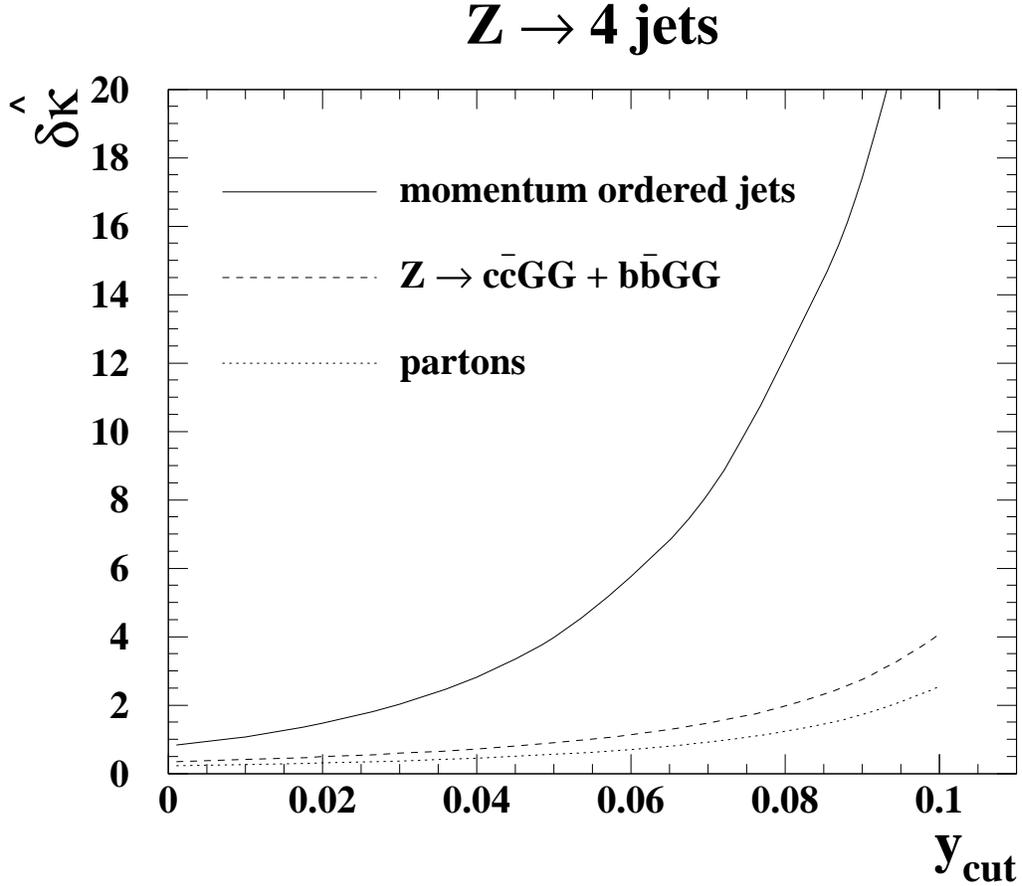}
    \parbox{0.9\textwidth}
{\caption{\it{
The error (inverse sensitivity) \dhg\ obtainable from the optimal
    observable as function of the jet resolution parameter $y_{cut}$
(\ref{jade}) assuming (\ref{nevents3g}) for the number of events. 
The solid curve
    represents the 
    results for momentum ordered jets (\ref{momorder}).
The dashed curve gives the sum of the results from reaction
    (\ref{proc4}) with $q'=c$ and $b$
    (cf. section~\ref{sec:partonscb3g}). The dotted curve
    represents the sum of the results from the reactions (\ref{proc4}) --
    (\ref{proc6}) under the assumption that one is able to
flavour-tag all partons and measure their momenta. 
    }}\label{fig:opthg}
}  \end{center}
\end{figure}

The contributions to the decay width are as in the parton case in
figure~\ref{fig:widhg}. Here suitable CP-odd tensor observables are
\begin{equation}
        T{'}_{ij}^{(n)} = (\widehat{\bf q}_1 - \widehat{\bf q}_2)_i \;
        (\widehat{\bf q}_1 \times 
\widehat{\bf q}_2)_j \; |\widehat{\bf q}_1 \times \widehat{\bf q}_2|^{n-2} + (i \leftrightarrow j) \; ,
\label{tenjet}
\end{equation}
where $n=1,2,3$ and $\widehat{\bf q}_l={\bf q}_l / |{\bf q}_l|$. As
expected from the results of section \ref{sec:partons3g} these are by far
less sensitive than
the respective optimal observable for momentum ordered jets. For this
reason, we investigate in the following only the optimal 
observable for momentum ordered jets.

We computed the inverse sensitivities \dhg\ for the optimal observable
(\ref{opthg}) for momentum ordered jets (\ref{momorder}) using (\ref{evopthg})
as a function 
of the jet resolution parameter \ycut\ (\ref{jade}). The results
are shown in figure~\ref{fig:opthg} (solid curve). The loss of information
about the parton charge and flavour
due to the ordering of the jets according the magnitude of their momenta leads
to a loss of sensitivity of the optimal observable. The difference between
the dashed curve (results from reaction (\ref{proc4}) with $q'=c$ and $b$)
and the dotted curve (results from the reactions (\ref{proc4}) --
    (\ref{proc6}) under the assumption that one is able to
flavour-tag all partons and measure their momenta) is essentially due to the
different number of events as one can see from tables~\ref{tab:optparthg}
-- \ref{tab:opthgbbgg}.

In table~\ref{tab:optjetshg}
in appendix~\ref{sec:numvalues} we list the coefficient of the
expectation value (\ref{evopthg}) for the optimal observable (\ref{opthg})
for momentum ordered jets (\ref{momorder}) for different values of the jet
resolution parameter $y_{cut}$ (\ref{jade}).

\section{Conclusions}

We have studied how one can search for CP violation in the 4 jet decays of
the Z boson assuming a CP-violating triple gluon coupling, which can
arise at one loop level in Higgs extensions of the SM \cite{dicus}, in
supersymmetric models \cite{tripgl2} or in left-right symmetric models
\cite{chang}.

We found that, for reasonable values of the coupling constants, the
additional contribution of the contact interaction to the 
decay width is at most at the per mille level. The decay width alone is
therefore not 
appropriate for determining the coupling constant.

We have investigated different tensor observables as well as the optimal
observables which can be used for the measurement of the
anomalous coupling. The tensor observables have only a low sensitivity on
the CP-violating coupling. No simple observables with sensitivities near to
the optimal have been found.

If it is possible to tag the flavour and the charge of all jets then, 
with a {\em total} number of $N_{tot}(\ycut=0.01) \cong 6 \cdot 10^6$ Z
decays and choosing a jet resolution parameter $y_{cut} = 0.02$ the
anomalous coupling can be determined with an accuracy of order 0.3 at
1~s.~d. level using the optimal observable. If flavour tagging is possible
for $b$ and $c$ quarks and taking together the subprocesses
(\ref{proc4}) with $q'=c$ and $q'=b$ an accuracy of order 0.5 at
1~s.~d. level is reachable.
If no flavour is tagged and instead all four jets are ordered according to
the magnitude of their momenta, the coupling constant \hg\ can be measured
with an accuracy of order 1.5 using the same total number of Z 
decays (see figure \ref{fig:opthg}). 

In our theoretical investigations we assumed always $100\%$ efficiencies
and considered the statistical errors only. But the total number of Z
decays collected by the LEP and SLC experiments together is of order
$10^7$. Thus the accuracies in the determinations of \hg\ discussed
above should indeed be within experimental reach.

As shown in \cite{tripgl4} the CP-violating triple gluon coupling can also
be studied in the reaction $ p \bar{p} \rightarrow 3\; {\rm jets}\, + X$, for
instance at the Tevatron. The accuracies obtainable there for \hg\ are
generally of similar order of magnitude as for Z decays (cf. eqs. (4.10)
and (4.12) of \cite{tripgl4}). In detail they depend, of course, on the
number of available events within cuts.

It is interesting to compare these accuracies to the limit on $|\hg|$ one
receives under the assumption that the triple gluon operator 
(\ref{lcp}) delivers the main contribution to the electric dipole moment (EDM)
$d_n$ of the neutron. At the moment the upper limit on 
$d_n$ is at $90\%$ c. l. \cite{pdg98}:
\begin{equation}
    d_n < 0.97 \cdot 10^{-25} \;e\,{\rm cm} \;\;.
\end{equation}
Using the ``naive dimensional analysis'' described in \cite{tripgl1}, but
using the correct anomalous dimension for the
3-gluon-operator (\ref{lcp}) as in \cite{tripgl3}, we get
\begin{equation}
   |\hg| < 2.2 \cdot 10^{-7} \;\;.
\end{equation}
This is much smaller than the accuracies reachable in our analyses with
presently available numbers of Z decays.
However, one should keep in mind, that many 
CP-odd operators can contribute to $d_n$ and cancellations among them cannot be
excluded. 

To summarize: We have shown that the measurement of the optimal observables for
flavour tagged and momentum ordered jets in 4 jet decays of the Z will give
useful limits on \hg.
(A FORTRAN-program for the optimal observables is available from the
authors.\footnote{World
Wide Web address: {\tt http://www.thphys.uni-heidelberg.de/$\:\widetilde{
    }\,$schwanen}}) 
Even if the limits obtainable are much worse than from the neutron's EDM
they have the advantage of being direct, i.~e. of involving explicitly 3
gluons in the splitting of one gluon into 2 gluon jets. Such studies
with the LEP1 data could also serve as pilot studies for future
investigations of this sort in the process $e^+ e^- \rightarrow$ 4 jets
at planned linear colliders (see e.~g. \cite{linac}), where due to the
higher c.~m. energy of 0.5 to 1 TeV effects of the dimension 6 effective 3
gluon coupling (\ref{lcp}) are of course enhanced.

\enlargethispage{\baselineskip}
\pagebreak
\subsection*{Acknowledgements}
We would like to thank W.~Bernreuther, A.~Brandenburg, S.~Dhamotharan,
M. Diehl, P.~Haberl, A.~Hebecker, W.~Kilian, J.~von Krogh, R.~Liebisch,
P.~Overmann,
S.~Schmitt, M.~Steiert, D.~Topaj and M.~Wunsch for valuable discussions.


\begin{appendix}

\section*{Appendix A \ \ Numerical Values}
\label{sec:numvalues}

We list some numerical results for the coefficient of the expectation value
for the optimal observable for partons and momentum ordered jets in the
final state. The statistical errors of the
numerical calculation are typically at the per cent level.

\begin{table}[H]
  \begin{center}
\begin{tabular}{|c||c|}
\hline
$y_{cut}$ & $c$ \\
\hline \hline
0.01    & $3.014\cdot 10^{-5}$\\ 
\hline      
0.02    & $6.158\cdot 10^{-5}$\\
\hline     
0.05    & $1.587\cdot 10^{-4}$\\
\hline                            
\hline                            
\end{tabular}
\parbox{0.9\textwidth}{\caption{\it{
      The numerical values of the coefficient of the expectation value
      (\ref{evopthg}) for the optimal observable $\OBS=O$
      (\ref{opthg}) for different values of the jet resolution parameter
      $y_{cut}$ (\ref{jade}) for partons in the final state
      (sum of the reactions (\ref{proc4}) -- (\ref{proc6})).
    \label{tab:optparthg}}}}
  \end{center}
\end{table}
%
\begin{table}[H]
  \begin{center}
\begin{tabular}{|c||c|}
\hline
$y_{cut}$ & $c$ \\
\hline \hline
0.01    & $3.172\cdot 10^{-5}$\\ 
\hline      
0.02    & $6.489\cdot 10^{-5}$\\
\hline     
0.05    & $1.693\cdot 10^{-4}$\\
\hline                            
\hline                            
\end{tabular}
\parbox{0.9\textwidth}{\caption{\it{
      The numerical values of the coefficient of the expectation value
      (\ref{evopthg}) for the optimal observable $\OBS=O$
      (\ref{opthg}) for different values of the jet resolution parameter
      $y_{cut}$ (\ref{jade}) from reaction \ZCG\ (\ref{proc4}).
    \label{tab:opthgccgg}}}}
  \end{center}
\end{table}
\begin{table}[H]
  \begin{center}
\begin{tabular}{|c||c|}
\hline
$y_{cut}$ & $c$ \\
\hline \hline
0.01    & $3.159\cdot 10^{-5}$\\ 
\hline      
0.02    & $6.515\cdot 10^{-5}$\\
\hline     
0.05    & $1.695\cdot 10^{-4}$\\
\hline                            
\hline                            
\end{tabular}
\parbox{0.9\textwidth}{\caption{\it{
      The numerical values of the coefficient of the expectation value
      (\ref{evopthg}) for the optimal observable $\OBS=O$
      (\ref{opthg}) for different values of the jet resolution parameter
      $y_{cut}$ (\ref{jade}) from reaction \ZBG\ (\ref{proc4}).
    \label{tab:opthgbbgg}}}}
  \end{center}
\end{table}
%
\begin{table}[H]
  \begin{center}
\begin{tabular}{|c||c|}
\hline
$y_{cut}$ & $c$ \\
\hline \hline
0.01    & $1.709\cdot 10^{-6}$\\ 
\hline      
0.02    & $2.666\cdot 10^{-6}$\\
\hline     
0.05    & $3.207\cdot 10^{-6}$\\
\hline                            
\hline                            
\end{tabular}
\parbox{0.9\textwidth}{\caption{\it{
      The numerical values of the coefficient of the expectation value
      (\ref{evopthg}) for the optimal observable $\OBS=O$
      (\ref{opthg}) for different values of the jet resolution parameter
      $y_{cut}$ (\ref{jade}) for momentum ordered jets
(chapter~\ref{sec:jets3g}).
    \label{tab:optjetshg}}}}
  \end{center}
\end{table}

\end{appendix}

\newpage


\thispagestyle{plain}
\pagestyle{empty}

\newpage

\addcontentsline{toc}{section}{\bf Literaturverzeichnis}

\begin{thebibliography}{99}

\bibitem{lepwg} The LEP Collaborations ALEPH, DELPHI, L3, OPAL, the 
  LEP Electroweak Working Group and the SLD Heavy Flavour and Electroweak
Groups: {\em A  
    Combination of Preliminary Electroweak Measurements and Constraints on
    the Standard Model}, CERN-EP/99-15. 

\bibitem{othertheo1} L. Stodolsky: Phys. Lett. {\bf B 150} (1985) 221;\\
  F. Hoogeveen, L. Stodolsky: Phys. Lett. {\bf B 212} (1988) 505.

\bibitem{othertheo2} J. F. Donoghue, B. R. Holstein, G. Valencia:
  Int. J. Mod. Phys. {\bf A 2} 
  (1987) 319;\\
  J. F. Donoghue, G. Valencia: Phys. Rev. Lett. {\bf 58} (1987) 451.

\bibitem{zdecay} W. Bernreuther, U. L\"ow, J. P. Ma, O. Nachtmann: Z. Phys.
  {\bf C 43} (1989) 117.

\bibitem{bernnach1} W. Bernreuther, O. Nachtmann: Phys. Rev. Lett. {\bf 63}
  (1989) 2787.

\bibitem{othertheo3}  J. Bernab\'{e}u, N. Rius: Phys. Lett. {\bf 232}
  (1989) 127;\\ 
  J. Bernab\'{e}u, N. Rius, A. Pich: Phys. Lett. {\bf 257} (1991) 219.

\bibitem{othertheo4} M. B. Gavela, F. Iddir, A. Le Yaouanc, L. Olivier,
  O. P\`{e}ne, 
  J. C. Raynal: Phys. Rev. {\bf D 39} (1989) 1870;\\
  A. De Rujula, M. B. Gavela, O. P\`{e}ne, F. J. Vegas: Nucl. Phys. {\bf B 357}
  (1991) 311.

\bibitem{othertheo5} S. Goozovat, C. A. Nelson: Phys. Lett. {\bf B 267}
  (1991) 128;\\ 
  Phys. Rev. {\bf D 44} (1991) 311.

\bibitem{xsec} J. K\"orner, J. P. Ma, R. M\"unch, O. Nachtmann, R. Sch\"opf:
  Z. Phys. {\bf C 49} (1991) 447.

\bibitem{ztotau} W. Bernreuther, G.W. Botz, O. Nachtmann, P. Overmann:
  Z. Phys. {\bf C 52} (1991) 567.

\bibitem{bernnach2} W. Bernreuther, O. Nachtmann: Phys. Lett. {\bf B 268}
  (1991) 424.

\bibitem{othertheo6}  G. Valencia, A. Soni: Phys. Lett. {\bf B 263} (1991)
  517.

\bibitem{bernnachov} W. Bernreuther, O. Nachtmann, P. Overmann:
  Phys. Rev. {\bf D 48} (1993) 78.

\bibitem{othertheo7} K. J. Abraham, B. Lampe: Phys. Lett. {\bf B 326}
  (1994) 175.

\bibitem{width} W. Bernreuther, G. W. Botz, D. Bru\ss, P. Haberl, 
  O. Nachtmann: Z. Phys. {\bf C 68} (1995) 73.

\bibitem{higgs} W. Bernreuther, A. Brandenburg, P. Haberl,
  O. Nachtmann: Phys. Lett. {\bf B 387} (1996) 155.

\bibitem{hab} P. Haberl: {\em CP Violating Couplings in Z $\rightarrow$ 3 Jet
  Decays Revisited}, hep-ph/9611430.
  
\bibitem{remarks} W. Bernreuther, O. Nachtmann: Z. Phys. {\bf C 73} (1997) 647.

\bibitem{over} D. Bru\ss, O. Nachtmann, P. Overmann:
  Eur. Phys. J. {\bf C 1} (1998) 191. 

\bibitem{opaltautau} P.D. Acton et al., (OPAL Collaboration):
  Phys. Lett. {\bf B 281} (1992) 405.

\bibitem{alephtautau92} D. Buskulic et al., (ALEPH Collaboration):
  Phys. Lett. {\bf B 297} (1992) 459.

\bibitem{alephtautau95} D. Buskulic et al., (ALEPH Collaboration):
  Phys. Lett. {\bf B 346} (1995) 371.

\bibitem{opaltautauopt} R. Akers et al., (OPAL Collaboration): Z. Phys. {\bf
    C 66} (1995) 31.

\bibitem{aleph} D. Buskulic et al., (ALEPH Collaboration):
  Phys. Lett. {\bf B 384} (1996) 365.



\bibitem{l3mumuga} M. Acciarri et al., (L3 Collaboration): Phys. Lett {\bf
    B 436} (1998) 428.

\bibitem{opaltautaulimit} K. Ackerstaff et al., (OPAL Collaboration):
  Z. Phys. {\bf C 74} (1997) 403. 

\bibitem{opal} M. Steiert: {\em Suche nach CP-verletzenden Effekten in
  hadronischen 3-Jet Ereignissen mit bottom Flavour $Z^0 \rightarrow b
  \bar{b} G$}, Diploma thesis, University
  of Heidelberg (unpublished); \\
  Rainer Liebisch: {\em Suche nach CP-verletzenden Effekten au{\ss}erhalb des
  Standardmodells im Zerfall $Z^0 \rightarrow b \bar{b} g$}, Diploma thesis,
  University of Heidelberg (unpublished).

\bibitem{paper} O.~Nachtmann, C.~Schwanenberger: Eur. Phys. J. {\bf C 9}
(1999) 565.


\bibitem{morozov} J. Morozov: Sov. J. Nucl. Phys. {\bf 40} (1984) 505.

\bibitem{buchmueller} W. Buchm\"uller, D. Wyler: Nucl. Phys. {\bf B 268}
  (1986) 621.

\bibitem{tripgl1} S. Weinberg: Phys. Rev. Lett. {\bf 63} (1989) 2333.
   
\bibitem{tripgl4} A. Brandenburg, J. P. Ma, R. M\"unch, O. Nachtmann: Z. Phys. {\bf C 51} (1991) 225.

\bibitem{tripgl2} J. Dai, H. Dykstra, R. G. Leigh, S. Paban,
D. A. Dicus: Phys. Lett. {\bf 237B} (1990) 216; \\
J. Dai, H. Dykstra:
Phys. Lett. {\bf 237B} (1990) 256.

\bibitem{ellisnat} J. Ellis: Nature {\bf 344} (1990) 197.

\bibitem{dicus} D. A. Dicus, Phys. Rev. {\bf D 41} (1990) 999.

\bibitem{chang} Darwin Chang, Chong Sheng Li, Tzu Chiang Yuan:
  Phys. Rev. {\bf D 42} (1990) 867.

\bibitem{tripgl3} E. Braaten, Chong Sheng Li, Tzu Chiang Yuan:
Phys. Rev. Lett. {\bf 64} (1990) 1709.

\bibitem{pdg} R. M. Barnett et al. (PDG): Phys. Rev. {\bf D 54}
  (1996) 1.
  
\bibitem{dok} C. Schwanenberger: {\em CP-Verletzung in den
4-Jet-Zerf\"allen des Z-Bosons}, Doctoral thesis, HD--THEP 99--18,
University of Heidelberg (1999). 

\bibitem{form} The Symbolic Manipulation Program FORM.
  By J.A.M. Vermaseren (KEK, Tsukuba). KEK-TH-326, Mar 1992. 20pp. 

\bibitem{M} P.~Overmann, private communication, M - Reference Manual,\\
  http://www.thphys.uni-heidelberg.de/$\:\widetilde{
    }\,$overmann/M.html.

\bibitem{vegas} VEGAS: An Adaptive Multidimensional Integration Program.
  By G.Peter Lepage (Cornell U., LNS). CLNS-80/447, Mar 1980. 30pp. 

\bibitem{jade} S. Bethke et al. (JADE Collaboration): Phys. Lett. {\bf B
    213} (1988) 235. 

\bibitem{comphep} P.A.Baikov et al., Physical Results by means of
  CompHEP, in Proc.of X Workshop on High Energy Physics and Quantum
  Field Theory (QFTHEP-95), ed.by B.Levtchenko, V.Savrin, Moscow:
  hep-ph/9701412 (1996) 101; \\ 
  E.E.Boos, M.N.Dubinin, V.A.Ilyin, A.E.Pukhov, V.I.Savrin:
  hep-ph/9503280.

\bibitem{opt} D. Atwood, A. Soni: Phys. Rev. {\bf D 45} (1992)
  2405; \\
  M. Davier, L. Duflot, F. Le Diberder, A. Roug\'{e}:
  Phys. Lett. {\bf B 306} (1993) 411.

\bibitem{opt1} M. Diehl, O. Nachtmann: Z. Phys. {\bf C 62} (1994) 397.

\bibitem{opt2} M. Diehl, O. Nachtmann: Eur. Phys. J. {\bf C 1}
  (1998) 177. 
     
\bibitem{pdg98} C. Caso et al. (PDG): Eur. Phys. J. {\bf C 3} (1998) 1.

\bibitem{linac} R. Brinkmann, G. Materlik, J. Rossbach, A. Wagner: {\em
Conceptual Design of a 500 GeV $e^+ e^-$- Linear Collider with Integrated
X-Ray Laser Facility, Vol. 1-2}, DESY-97-048, ECFA-97-182;\\
{\em $e^+ e^-$ Collisions at 500 GeV;
The Physics Potential, Part A -- D}, Ed. P.~M.~Zerwas, Reports DESY-92-123A,
B (1992), DESY-93-123C (1993), DESY-96-123D (1996).

\end{thebibliography}
\end{document}